# CommuniSense: Crowdsourcing Road Hazards in Nairobi

Darshan Santani[1,2], Jidraph Njuguna[3], Tierra Bills[4], Aisha W. Bryant[4], Reginald Bryant[4], Jonathan Ledgard[2] and Daniel Gatica-Perez[1,2]

[1]Idiap Research Institute, Switzerland
[2]Ecole Polytechnique Fédérale de Lausanne (EPFL), Switzerland
[3]Florida State University, USA
[4]IBM Research Africa, Kenya

dsantani@idiap.ch, jn08f@my.fsu.edu, tbills@ke.ibm.com, awalcott@ke.ibm.com,
bryantre@ke.ibm.com, jonathan.ledgard@epfl.ch, gatica@idiap.ch

## ABSTRACT
Nairobi is one of the fastest growing metropolitan cities and a major business and technology powerhouse in Africa. However, Nairobi currently lacks monitoring technologies to obtain reliable data on traffic and road infrastructure conditions. In this paper, we investigate the use of mobile crowdsourcing as means to gather and document Nairobi's road quality information. We first present the key findings of a city-wide road quality survey about the perception of existing road quality conditions in Nairobi. Based on the survey's findings, we then developed a mobile crowdsourcing application, called *CommuniSense*, to collect road quality data. The application serves as a tool for users to locate, describe, and photograph road hazards. We tested our application through a two-week field study amongst 30 participants to document various forms of road hazards from different areas in Nairobi. To verify the authenticity of user-contributed reports from our field study, we proposed to use online crowdsourcing using Amazon's Mechanical Turk (MTurk) to verify whether submitted reports indeed depict road hazards. We found 92% of user-submitted reports to match the MTurkers judgements. While our prototype was designed and tested on a specific city, our methodology is applicable to other developing cities.

## ACM Classification Keywords
H.4.m [**Information Systems Applications**]: Miscellaneous

## INTRODUCTION
In the last decade, Nairobi (population of 3.1 million in 2009) has experienced rapid urbanization, which has led to a rise in traffic congestion and long commute times. This has resulted in growing frustration amongst commuters [23]. While there has been significant growth in car ownership and informal bus transit (known as *matatus*), the transportation infrastructure has not kept pace with this growth. It is estimated that traffic congestion in Nairobi costs the economy an estimated 37 billion Kenyan Shillings annually (equivalent to 413 million USD) [7]. This trend is unsustainable and detrimental to the achievement of Kenya's 2030 development plans [40].

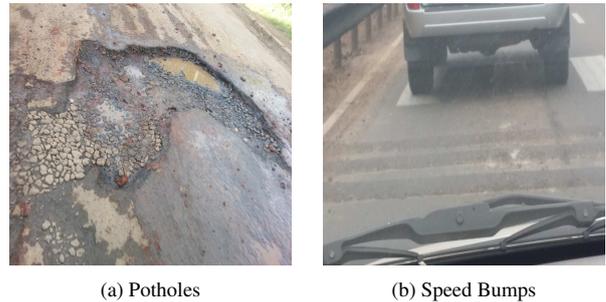

(a) Potholes     (b) Speed Bumps

Figure 1: Road surface conditions in Nairobi (a) Potholes (b) Speed Bumps.

In addition to this growth in travel demand, Nairobi has not received adequate attention with regard to long term transportation planning [20]. Nairobi roads are known for their hazardous conditions, which include gaping potholes, unregulated speed-bumps and abrupt road surface changes. Figure 1 shows some of the road hazards on the streets of Nairobi. In the rest of the paper, by "road hazards", we specifically refer to potholes and speed-bumps. Although speed-bumps are traditionally used for traffic calming and speed mitigation, in Nairobi they are frequently unlabeled, poorly (and often irrationally) placed, and are not accompanied with proper signage. For example, in one of our field tests on a 2.4km stretch of road where the speed limit is 60km/h, under free flow conditions we encountered 13 speed-bumps resulting in an average travel speed of 20km/h.

In the field of mobile sensing, there has been research interest to automatically detect potholes and monitor road surface conditions using mobile sensor data [19, 32, 26, 37]. In the *Pothole Patrol*, the authors presented a machine learning based approach to detect potholes, using accelerometer and GPS data [19]. Mednis et al. [26] described a pothole detection algorithm, employing accelerometer data obtained using Android based smartphones. In contrast with the mobile sensing domain, the transportation research community has proposed road surface monitoring using camera-based systems [25]. Most of the systems and algorithms described above rely on manually collected ground truth data which serve as training data. Collecting training data this way, requires careful planning and experimentation, which typically involves repeatedly driving a set of road segments and manually labeling the location of potholes and other forms of road anomalies. Relying on hand-labeled datasets severely limits the scale, spatial coverage and amount of data available to train classifiers. Therefore, we believe that mobile crowdsourcing provides an alternate scalable solution to collect labeled data for developing future automated platforms.

In intelligent transportation systems (ITS) and traffic research domain, the use of crowdsourcing methods have began to receive attention [12, 24, 31]. In *CrowdITS*, the authors proposed a hybrid system to integrate crowd-based reporting with GPS-based navigation system, to suggest congestion-free routes [12]. In [24], the author advocated the use of mobile social media and collaborative applications to increase social interactions on the road. Crowdsourcing in ITS present exciting opportunities for developing cities, as they lack monitoring technologies to obtain reliable data on traffic and road infrastructure conditions. The costs associated with deploying sensing infrastructure to monitor road quality in developing urban areas are often prohibitive, and impractical to implement; therefore it becomes imperative to leverage locally available resources to collect this information. In this paper, we examine the use of mobile crowdsourcing as means to obtain road infrastructure data in large developing cities, particularly Nairobi. Our work addresses the following two research questions:

**RQ1:** What are the perceptions of citizens in large developing cities towards the state of existing road infrastructure conditions?

**RQ2:** How can mobile crowdsourcing technology be leveraged to support citizen-based data collection and verification of road infrastructure conditions in developing cities, like Nairobi?

A crowdsourcing approach, using smartphones, is promising due to the widespread penetration of mobile devices (78.2% mobile penetration in Kenya in 2013 [29]) and the increasing popularity of smartphones. Smartphone penetration has been fuelled by the introduction of low-cost Android phones and the trend is expected to continue as different vendors including Google, Huawei, and LG plan to roll out more low-cost smartphone devices in Kenya [39, 10]. This trend is similarly observed in other developing cities as well.

In this paper, we present a prototype system to address the problem of documenting Nairobi's road infrastructure conditions. We first designed a travel survey to understand the existing state of road quality conditions in this city. The survey questionnaire collected demographics, weekly travel practices, perception of current road quality conditions and impact on their travel experience. To account for socio-economic bias, we conducted the survey via two different channels.

Based on the survey's findings, we then developed a mobile crowdsourcing application, called *CommuniSense* to collect data on road surface conditions. The application allows users to submit road hazard reports where they locate, describe, and take pictures of road hazards. We test our application through a two-week field study amongst 30 participants, who submitted a total of 254 reports characterizing various forms of road hazards from different areas in Nairobi. To verify the authenticity of user-contributed reports from our field study, we propose to use online crowdsourcing using Amazon's Mechanical Turk (MTurk), to verify whether submitted reports indeed depict road hazards. We found 92% of user-submitted reports to match the MTurkers judgements. While our prototype was designed and tested on a specific city, our methodology is equally applicable to other developing cities.

Integrating the collection of mobile sensor data (as done in mobile sensing and ITS domain) with crowdsourced data on road infrastructure conditions (as proposed in *CommuniSense*), our broader objective is to build a travel model to estimate travel speeds, fuel consumption, and vehicle emissions, as a function of road infrastructure conditions. Furthermore, we envision *CommuniSense* as a system to facilitate citizen engagement and participation for small-scale community infrastructure maintenance activities.

**RELATED WORK**

In the developing world, one of the most common ways to collect data is via text messages or SMS. The low cost of feature phones and wide availability of SMS service has enabled various SMS-based data collection systems including RapidSMS [3], FrontlineSMS [14], and Ushahidi [30]. FrontlineSMS has been designed to gather unstructured data, while RapidSMS has been designed primarily for structured data. The Ushahidi platform extended FrontlineSMS and was deployed first in Kenya during the 2007 post-election violence. The platform allowed Kenyans to submit violence related reports using SMS (and email).

Despite the popularity of SMS, SMS-based tools are often unreliable and expensive. The costs associated with sending 1Mb data over SMS is over 3600 times more expensive than GPRS (General Packet Radio Service) [11]. Furthermore, SMS-based tools cannot provide fine-grain location and high-quality image data. Although these platforms have been successful deployed in the past, they provide a bare minimum support for user interactivity and are designed to be deployed in environments experiencing financial, social, political, or natural disaster hardships.

In the recent past, the increasing popularity of smartphones and increasing investment in cellular infrastructure has generated excitement for smartphone-based crowdsourcing solutions in developing regions. This growth has provided major opportunities to collect data in a cost effective manner. Tools like OpenDataKit (ODK) [22] and Nokia Data Gathering [2], have been designed primarily for the developing world. ODK is a smartphone-based platform designed to build data collection solutions for organizations with limited financial and technical resources (e.g., NGOs). In a more recent work, the team behind ODK, redesigned its architecture to simplify the process of creating and managing data collection pipelines for individuals with limited technical experience [18].

Technically, *CommuniSense* is similar to ODK 2.0. We designed and build our system from scratch to integrate incentive (financial/social) [21, 41], gamification, crowdsourced verification and social media modules for future needs. To the best of our understanding, incorporating these modules in ODK would require systemic changes to its core architecture.

## ROAD QUALITY SURVEY

Most surveys in Kenya have focused on either traffic congestion [40], travel choice behavior [34] or mobile penetration and usage [13]. To the best of our knowledge, no digitized survey has been conducted to understand the opinions of people on the state of road quality in the context of Nairobi or Kenya in general. We conducted a travel survey in Nairobi with two goals. First, we wanted to understand what Nairobi travelers think of the existing state of road quality in their city. Second, we wanted to gage their willingness to engage and participate in reporting information on road hazards to support government in urban road maintenance. The survey questionnaire had a series of questions asking respondents about themselves, how they travel on a weekly basis, and how they rate the current road quality conditions based on their daily travel experience. Specifically, our survey had four themes:

- **Mode of transport**: In this section, we asked respondents about their frequency of usage of different transportation modes on a weekly basis. We focused on four transport mode choices: personal vehicle, matatus or bus, taxis, and walking. Matatus are privately-owned informal minibuses that form the backbone of transportation network in Nairobi.

- **Status quo on road quality**: In this section, we explored the current state of road quality in respondents' residential neighborhood, workplace neighborhood, and Nairobi at large, on a five-point scale ranging from *very poor* (1) to *very good* (5). We also asked participants to rate potholes and speed bumps as major road nuisances on a five-point Likert scale ranging from *strongly disagree* (1) to *strongly agree* (5).

- **Overall impact of road hazards**: In this section, we asked users about the impact of road hazards on their travel discomfort and their personal vehicle's wear and tear (if they owned a personal vehicle). In the survey, we used "road hazards" as an umbrella term to refer to potholes, speed bumps, cracks on the road surface, abrupt pavement changes, or uneven road surface conditions; we made this definition explicit to the respondents.

- **Reporting road hazards**: In this section we quizzed users on their knowledge about how to report a road hazard to the city council, and if they had reported any in the past in this way. In addition, we asked them about their preferred choices and motivations for reporting road hazards.

- **Demographics**: We asked participants about their demographic characteristics (age and gender), living status, and whether they own an Android-based smartphone.

The majority of the survey questions were multiple choice where respondents chose from a list of options. In addition, we had two open-ended questions where respondents were asked about the name of their residential and workplace neighborhood (as free-form text). All the multiple-choice questions were mandatory, while the open-ended questions were optional. In total, the survey consisted of 18 mandatory and 2 optional questions. Responses were anonymous.

For conducting the survey, we used two different channels: web-based (online) and SMS-based (offline). We used two different channels to account for any potential demographics bias. On one hand, we believe that an online survey would typically target upper class, upper middle class, and expatriate communities, while on the other hand a SMS-based survey would cater more to working class and non-smartphone users who typically do not have easy access to the internet. We acknowledge that our surveys are not representative of population of Nairobi as no stratification technique or demographic sampling was applied while selecting users.

### Online Survey (gSurvey)

In this channel, we used an online platform (Google Forms) to conduct the survey. The survey was distributed via email to mostly university students, and internally within our organization. In addition to asking the respondents to complete the survey, we also ask them to share the survey on various social media channels (including Twitter and Facebook) to reach a larger audience. We also posted the survey on our organization's Twitter and Facebook official pages. No monetary incentives were provided for answering the survey.

### SMS-based Survey (mSurvey)

Our second distribution channel was a SMS-based mobile survey platform using mSurvey [1]. mSurvey is a Nairobi-based company which provides a mobile platform to conduct surveys and market research in Kenya. In order to have a wider reach and specifically address a population that does not have easy access to the internet, we used mSurvey's platform, where respondents receive each question per SMS on their mobile devices.

The mSurvey platform used a sample of 500 users randomly selected from their existing worker population in Nairobi. No stratification techniques and demographics filters were applied while selecting the user sample in our survey. Respondents received 40 Kenyan Shillings (equivalent of 0.5 USD) as financial reward to complete the full survey.

### Results

For the online survey (gSurvey), we received a total of 442 responses, while for the SMS-based survey (mSurvey), we received a total of 439 completed responses. In total we have a pool of 881 respondents to our survey. In this section we describe the results of both surveys. We focus on five survey themes, which are relevant to the scope of our work.

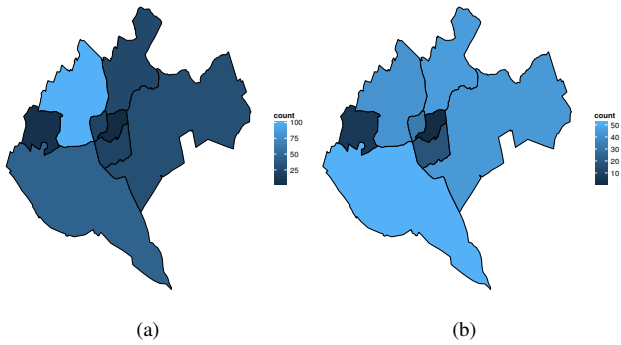

Figure 2: Nairobi residential neighborhoods of respondents for a) web-based survey, and b) SMS-based survey. Spatial information of different administrative areas and neighborhoods of Nairobi are obtained via [5].

*Demographics*
In the online survey, 62% of respondents were male, and 37% of them were female, while the remaining participants chose not to share their gender identity. On the other hand, amongst the mSurvey participants, 58% of respondents were male, and 42% of them were female. For the age distribution, we observe that amongst the gSurvey (resp. mSurvey) population, 23% (resp. 66%) belong to 18–24 age group, 47% (resp. 26%) belonged to the age category of 25–34 years, and 25% (resp. 7%) belong to 35–50 age segment. For both surveys, we did not had a single participant below 18 years old. From these results, it is clear that both surveys cater to different population demographics. In terms of smartphone ownership, 76% of gSurvey respondents owned an Android-based smartphone, while for the mSurvey population 50% of them owned one.

As mentioned in the previous paragraph, participants were asked to list the name of their residential and workplace neighborhood. Of all the users who provided a response, we geocoded their neighborhood addresses to geographic co-ordinates (latitude and longitude pairs). Figure 2 shows the spatial coverage of participants' residential neighborhoods in Nairobi, for both surveys. Based on the local knowledge of the city, we observe that high-income neighborhoods (like Karen, Kilimani and Kileleshwa) are represented more in the web-based survey, when compared to SMS-based survey.

*Status quo on road quality*
In the online survey, 47% (resp. 30%) of respondents rated road quality as either *poor* or *very poor* in their residential (resp. workplace) neighborhoods (see Figure 3a). Consistent with the online survey, the majority of mSurvey respondents 55% (resp. 38%) rated the quality of roads in their residential (resp. workplace) neighborhoods, as either *poor* or *very poor*. Figure 3a shows the distribution of road quality in residential neighborhoods across the entire response scale, which clearly highlights that both survey populations find the state of road quality at places where they live to be dismal, with a more negative perception for the mSurvey participants.

When survey takers were asked to rate the road quality in Nairobi at large (i.e., not only for home and work neighborhoods), 45% of online respondents found the existing road surface conditions to be bad (*poor* or *very poor*). Surprisingly, only 20% of SMS-based respondents considered the overall Nairobi roads to be in bad condition, with an overwhelming 42% found the roads in good shape (*good* or *very good*). This is in contrast to their perception of their personal neighborhoods discussed in the previous paragraph. There might be some aspirational factors at play here; this would have to be investigated in future work.

Of all the online respondents, 79% *agreed* or *strongly agreed* that potholes are a major road nuisance, while 67% of the mSurvey population acknowledged this fact. Figure 3b compares this trend across both populations and the entire scale. 42% of web-based and 29% SMS-based respondents *agreed* or *strongly agreed* that speed-bumps are a major road inconvenience. These findings substantiate our intuition that potholes and speed-bumps are indeed perceived as road hazards, with the SMS population being less sensitive to this issue.

*Impact of road hazards*
While considering the impact of road hazards, 65% of gSurvey and 46% of mSurvey respondents considered road hazards to cause either *major* or *severe* impact on their personal travel comfort, as shown in Figure 3c. Of all the online survey takers who own a personal vehicle, 77% of people considered road hazards to have a *major* or *severe* impact on their vehicle's wear and tear. Note that while asking survey questions in this category, we explicitly defined "road hazards" as potholes, speed bumps, road surface cracks, abrupt pavement changes, or uneven road surface conditions.

*Reporting road hazards*
Amongst the gSurvey population, 96% of respondents did not know the process of reporting the road hazard to Nairobi's city council. For the 4% who were aware of the process, 55% have ever reported information on a road hazard to the local administration. In contrast, 23% of SMS-based population were aware of the hazard reporting process, and 59% of them have reported one or more of these road hazard complaints.

Furthermore, in this category we also asked respondents about their preferred choice to report road hazards in Nairobi, even if they had never done it before. While asking this question in the online survey, users were given the freedom to choose more than one option as their response. We found that 70% of respondents chose mobile application as their preferred choice, while the second option was to report hazards via social media channels such as Twitter and Facebook (56%).

In contrast, for the SMS-based survey, we formulated this question as a ranking question, where users were asked to rank their top-3 preferred choices. Due to the lack of ranking feature in Google Forms, we formulated this question in the online survey as a multiple choice question with multiple responses. For their top preferred choice, only 4% and 26% of respondents chose mobile application and social media respectively, while 43% of respondents choosing a personal visit to the city council as their top choice to report hazards in Nairobi.

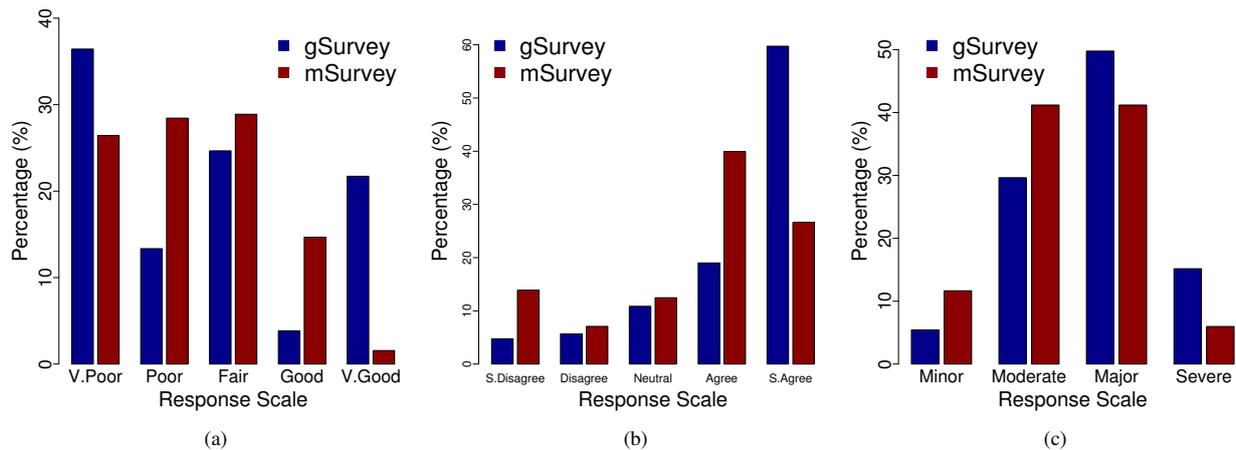

Figure 3: Plots showing the histograms for a) Road quality in residential neighborhood, b) Potholes as a major road nuisance, and c) Impact of road hazards on personal travel discomfort, for both online and SMS-based surveys.

These results point towards a clear difference in the populations; the SMS respondents are probably making their choices based on their most common interaction practice with government (face-to-face) in combination with lower degrees of mobile internet connectivity.

*Free-form user comments and feedback*

While conducting the online survey, respondents were given the option to voice their opinion and leave comments towards the end of the survey. It was not possible to provide that option to the mSurvey respondents due to the SMS inherent nature on how the survey was conducted. Out of 442 gSurvey participants, 101 left feedback encompassing different topics. We used topic modeling to perform basic content analysis and discover underlying topics from user comments. We used Latent Dirichlet Allocation (LDA) model [16], where each user comment was treated as a single document. After experimenting with different model parameters, the resultant topics did not provide any clear insights due to data scarcity.

As an alternative, we manually coded each comment in order to reveal common concepts and themes. We found that comments varied from general praise for the survey; personal commentary on the current state of traffic situation in Nairobi; negligence and the lack of any hope for a visible feedback from the local city council; and general advice on how to improve the existing situation. In the words of few users, here are some comments we find insightful:

*"Speed bumps are fine as long as they are marked so you don't just "discover" them with your head on the ceiling and stuff flying in the car. Look at the roads around the hospitals. I am sure a patient is half killed before they even get to the hospital for treatment ..."*

*"The state of our roads is dismal at best. Networks that were designed for a 90's population are being used, unchanged in the second decade of the 21st century"*

*"Road hazards is a major cause of road accidents in Kenya that should be addressed."*

*"Good job, keep it up. I look forward to seeing a site where I can report hazards and visualize whether the report has been received or not by the relevant authorities, and track of whether reported hazards are being fixed or not."*

## MOBILE APPLICATION

In developing cities, the lack of reliable infrastructure, limited connectivity, and inadequate resources make data collection difficult. Paper-based systems are a perennial favorite for city and government administrations. In Nairobi, the city council relies on these systems to handle road quality related complaints as well. However, the reasons that make paper popular are also its liabilities. These paper-based reporting systems lack transparency, accountability, and the speed at which reports are handled is very slow, leaving residents frustrated.

As part of our research, we visited the Nairobi city council engineering offices to learn more about the current reporting system. We found out that residents can use three options to report road hazards: phone calls, postal letters, or walk-in reports. Once the report is submitted, they are sent out to the engineering department for assessment. The engineer goes to the field to assess the hazard reported, takes pictures, documents exact location and severity, and determines how to best fix the hazard. When the engineer gets back to the office, they file a request for supplies which takes time to be fulfilled. This process usually lasts on average between three and six months and in some cases longer before appropriate actions are taken. We designed our crowdsourcing solution to improve this current process.

*CommuniSense* is a mobile crowdsourcing application, build on the Android platform, that is designed to collect data on road surface conditions in Nairobi (Figure 4). It is a relatively low-cost solution that leverages mobile technology. We chose to build *CommuniSense* on Android since it is cost effective, provides rich programmable interface, offers in-built graphics support and is supported across multiple devices. Using the Android platform, we can collect rich data including multimedia (images), location (GPS) and a myriad of other sensor data (accelerometer, Bluetooth, WiFi, etc.) which is not possible via a typical feature phone. Furthermore, the Android platform automatically optimizes the user interface (UI) experience on each device, while allowing as much control of the UI on different mobile device types.

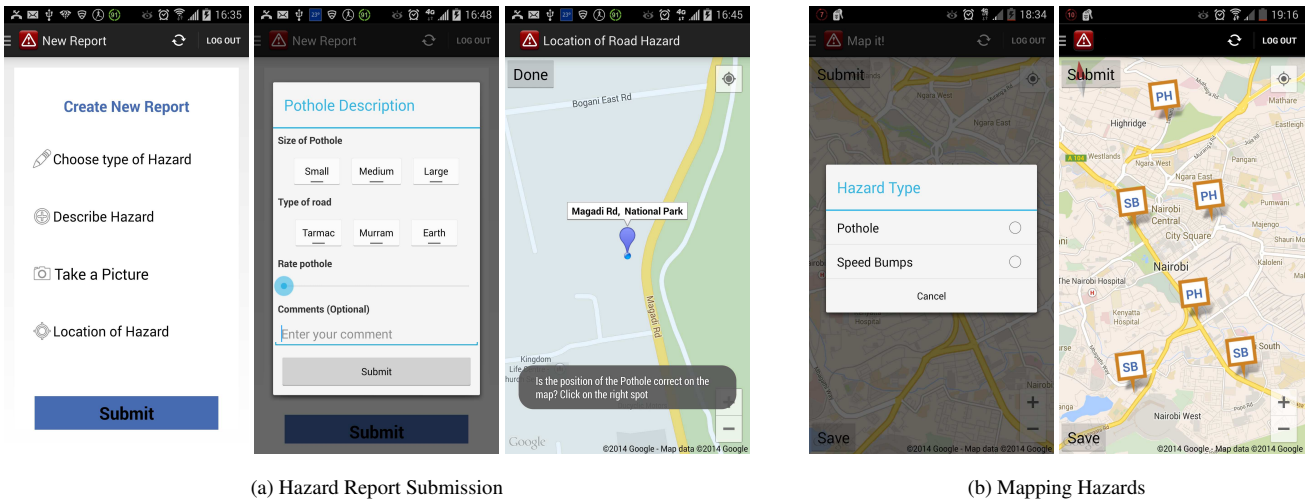

(a) Hazard Report Submission  (b) Mapping Hazards

Figure 4: Screenshots of the mobile app showing the sequence of stages for (a) Hazard report submission, and (b) Mapping hazards (in the map interface, PH stands for Potholes and SB stands for speed-bumps.)

Crowdsourcing the execution of microtasks to a diverse group of people offers unique advantages when combined with a highly motivated pool of workers. From the previously discussed survey, we found that the citizens wants to be engaged and are willing to participate in data collection. As per a travel survey conducted in [20], the mode of daily commutes in Nairobi is 47% by walking, 29% by matatu or mini-bus, 15% by private auto, 7% by other buses or shuttles, and 1% by other modes of transportation. This distribution avails a large segment of commuters that can provide manual reporting via *CommuniSense*. We are well aware that our smartphone-based approach presents constraints to our collection methodology in terms of reaching a much wider audience. However, recent trends in smartphone penetration and subsidized cost of smartphones in Kenya, demonstrate that they can be used to achieve sufficient data diversity [8, 39, 10].

This data collection platform provides us with hand-labeled hazard locations for two purposes. First, the geo-referenced images are valuable to document road hazards when displayed on a map. Second, we plan to use the geo-located data as training data for future work to detect and locate hazards using other phone sensor data, as done in other recent work [19, 27, 15]. The mobile application provides users with two reporting options which are described below.

**Hazard Report Submission**
In this option, users can submit a completely documented report which includes the type of road hazard, its description (including hazard's severity and road type), a picture showing the hazard, and its corresponding location (Figure 4a). To capture the location of the hazard, GPS sensor is triggered as soon as the user starts the application, so when the user is submitting the report, we automatically capture hazard's location.

While GPS provides accurate location estimates in the order of few meters, it suffers from few limitations, including urban canyon errors due to bad radio reception in areas surrounded by tall buildings (applicable in Nairobi downtown) and cold-start problems which result in inaccurate location estimates when a device is initially switched on. (During initial field tests, we found that the GPS coordinates in urban areas were in most cases off by 100–250m.) As a result, the mobile application gives users the functionality to update the location of the hazard (relative to its GPS-inferred location) by clicking and dragging the marker on the map, as shown in Figure 4a.

In Nairobi, mobile data is relatively expensive, and in remote areas the signal strength is weak to sustain a reliable data connection. Consequently, when the user is done completing the report they have two upload choices. Users can immediately submit the reports or save them locally on the device and upload them later, when reliable mobile data connection and/or access to Wi-Fi are available. Note that if the user decides to submit from the field and the data fails to get to the servers, it is automatically saved locally.

The reason to provide the offline reporting functionality was motivated by an initial discussion with a small set of commuters who raised concerns over the cost associated with uploading an image from the field. As a result, we perform image compression as a second mean, to reduce the costs associated with transferring the hazard report. We use Android's in-built `base64` encoding for image compression. When a user takes an image, we compress the image locally on the device and then send the compressed image to the backend for storage. While compressing images, we take into account the orientation of the mobile device (portrait or landscape mode) and its native resolution.

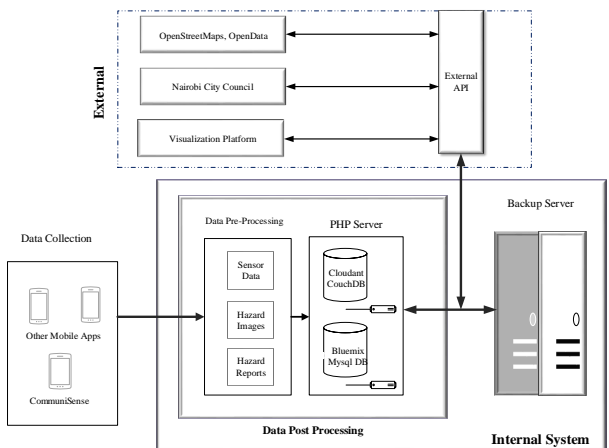

Figure 5: CommuniSense Backend Architecture

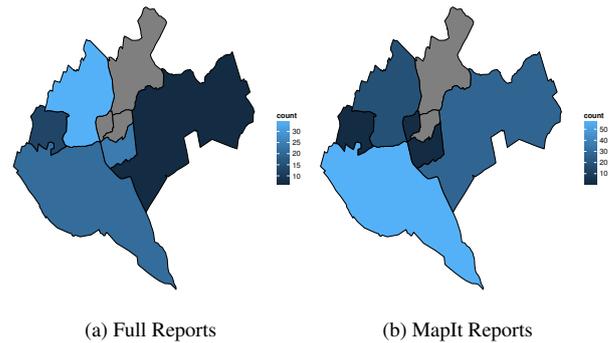

(a) Full Reports  (b) MapIt Reports

Figure 6: Spatial coverage of user-contributed submissions using *CommuniSense*: a) Full report submissions, b) Mapping hazards submissions. Regions colored gray do not have any user contributed submissions. The city division is the same as in Figure 2.

**Mapping Hazards (MapIt)**
The second option provides users with a quick way to report the location of a road hazard. Users are shown a map interface centered and zoomed to their current location (inferred from GPS). A click on the map prompts a dialog with a list of different hazards. Once the user selects a hazard name, a marker is placed on the map. As with the complete reporting, the user can adjust the location by clicking and dragging the marker. Figure 4b shows the data capturing process.

The idea behind this option is to give users the flexibility to report road hazards who are unable to fully document them during their commute. Drivers and commuters typically have intimate knowledge of the routes they frequently take and therefore can offer insights on the road surface conditions at a later time. This is a way to utilize the local knowledge of people to document road quality conditions in areas they are most familiar with, without requiring them to go through the process of submitting a complete report. The data collected using this process will be used in the future to build a probabilistic model to grade the quality of a road link.

**Backend Architecture**
The application is linked to a cloud `PHP` server that handles user authentication, receives reports, and handles all device-server communications (Figure 5). Users are required to create an account using their phone number. We anonymized the phone number and other identifying information to maintain users privacy. A unique `userID` is generated and associated with any activity between a user and the server. The hazard metadata (location, description, etc.) is saved in a `MySQL` database with images being saved as binary large objects.

**DATA COLLECTION EXPERIMENT**
To test the functionalities of *CommuniSense*, we conducted a two-week pilot user study. We performed a limited release of our mobile application to a selected number of participants. The application was published on Google Play Store but it was not available to everyone, as we wanted to test the functionality of the app with a limited set of users, before making it open for everyone. We control the access to our app via a private Google Plus community. Only users who were part of the community had access to the *CommuniSense* application.

To enroll participants in our study, we emailed 150 users, mostly college students from local universities and visiting students. Once a participant showed their willingness to be part of the study, we invited the participant to join our Google Plus community. After joining our community, participants had access to download and install the application on their devices. To motivate the users, we promised to award 500 Ksh (equivalent of 5.5 USD) to the top five contributors towards the end of the study. The top contributors were chosen based on the maximum number of legitimate and unique reports covering different neighborhoods in Nairobi.

**Results**
During our field experiment, of the 150 email invites sent, we had a total of 41 users who accepted our invitation to join the Google Plus community. Out of those 41 users, 30 installed the application (20% response rate).

During the two weeks of the trial, we had a total of 101 full report submissions and 153 MapIt submissions. Of all the full reports, 62% submission were of potholes, and the remaining 38% were of speed-bumps. Of all the MapIt submissions, 42% submission were of potholes, and the remaining 58.17% were of speed-bumps.

Out of 101 full reports submitted, 99 of them came from Nairobi county (61 potholes and 38 speed-bumps), while for the MapIt submissions, 109 came from Nairobi county. Figure 6 shows the spatial coverage of the reports from within Nairobi city limits. One can observe that there were at least a few reports from most neighborhoods. although some of the regions are missing from our field experiment.

Of the 61 full reports of potholes submitted, we observe that 43% of potholes were rated *minor*, and 31% were rated either *major* or *severe* on the severity scale, as shown in Figure 7a. Note that field users rated the severity of potholes on a 4-point scale ranging from *minor* (1) to *severe* (4). When observing the speed-bumps, we note that 55% of field users encountered an unlabeled speed-bump (Figure 7b).

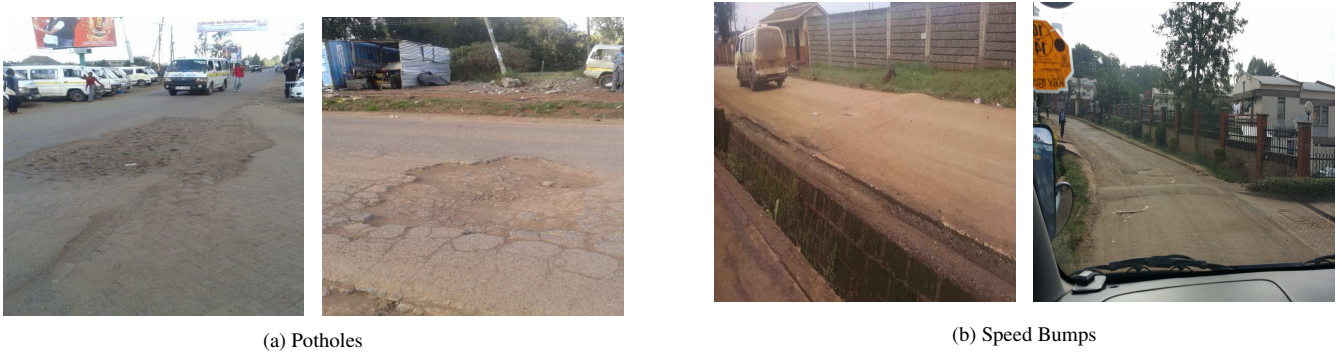

(a) Potholes    (b) Speed Bumps

Figure 8: Sample images from the field study. Two random images from our dataset reported as (a) Potholes, and (b) Speed Bumps. Note that images showing faces and license plate numbers have been blurred or masked.

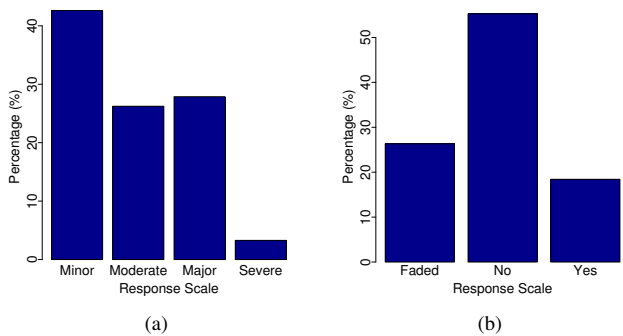

Figure 7: Histogram showing hazard attributes for (a) Severity of potholes, and (b) Labeling of speed-bumps, as reported by field users.

**IMAGE VERIFICATION EXPERIMENT**

Crowdsourcing offers opportunities for people to supplement their income in developing countries. However, the openness of access to crowdsourcing platforms often leads to malicious and spam behavior, and sometimes sabotage. As an example, for the well-known DARPA network challenge, the winning entry received 80% of malicious submissions [38]. In another example, Ushahidi, the crowdsourcing platform for social activism and crisis mapping, shut down their operations during the 2011 Arab spring due to growing concern that governments official might use the platform to track the activities of people [28].

For most of the tasks available on crowdsourcing services like Mechanical Turk, Crowdflower, or MobileWorks, financial incentives need to be in place to motivate the worker population to participate. However, when the crowdsourcing task involves monetary incentives, users might put in only minimum effort to secure the financial reward. As a result, quality control in crowdsourced data cannot be neglected. This presents several research challenges.

In our case, we have to verify that a) a contributed submission indeed depicts the road hazard as reported by the user, and b) it is located at the claimed location. The second verification task is conceptually feasible as the location information is automatically captured via the GPS location sensor most times. However, the authenticity of the reported road hazard and its details is more subjective to evaluate. We present a crowdsourced approach for this in the next subsection.

**Crowdsourcing Image Verification**

We designed and conducted a crowdsourcing study to assess whether the images obtained via our mobile application can display road hazard properties. For crowdsourcing, we used MTurk and chose US-based workers with at least 95% approval rate for historical HITs (Human Intelligence Tasks). In addition, to increase the potential reliability of MTurk annotations, we only chose "Master" annotators, which typically involves a worker pool with an excellent track record of completing thousands of tasks with precision.

For each HIT annotation task, the annotators were shown one image and asked to classify the image as either pothole or speed-bump. Given the annotators' choice, they were further asked to describe the chosen hazard. If the user categorized an image as a pothole, users were asked to further describe it in terms of size and severity. For size, users were given the option to choose from *small*, *medium* and *large*; for severity, users were given a four-point scale ranging from *minor* (1), *moderate* (2), *major* (3) and *severe* (4). If the user chose speed-bump as the option we asked them to describe its size, number of bumps, and whether the speed-bump was labeled or painted. If the user was unable to classify an image as either containing a pothole or a speed-bump, then an option was given to mark whether the image contained both a pothole and a speed-bump; or showed uneven or cracked road surface; or the image showed a smooth road surface. For the MTurk experiment, we randomly chose 50 images from the set collected in the field experiment. We collected 10 different annotations for each image. Consequently, we collected a total of 500 responses for every question. Every worker was reimbursed 0.15 USD per HIT (i.e., per image)

The questions asked to describe the hazard were identical to the questions shown to users while reporting from the wild i.e., using our mobile application. For these questions, no explicit definitions of a pothole or speed-bump were provided, so workers needed to rely on their internal representation. All the images shown to the users were anonymized. To the best of our ability, we avoided images where one can potentially identify faces or skin color, to protect the privacy of individuals and reduce any potential bias while characterizing the road quality. Moreover, we ensured that images that showed the license plate numbers or any other information that could explicitly reveal the identity of the city under study were masked e.g., an image showing street banners with the word Nairobi in it. Image examples are shown in Figure 8.

| Method | Potholes | Speed Bumps |
|---|---|---|
| Majority Voting | 34 (100%) | 12 (75%) |
| Median | 34 (100%) | 12 (75%) |

Table 1: Table showing summary statistics for aggregation methods. For each method, we show the total number and percentage (shown in brackets) of correctly classified images for both road hazards (i.e., where the consensus between the MTurk population and the field experiment matches)

### Results
In this section we present the results of our image verification experiment.

*Completion Rate*
For the MTurk experiment, we had a pool of 39 workers who responded to our HITs. For a total number of 500 HIT assignments available in this experiment, we observe that a typical worker completed an average of 13 HITs, while they could potentially undertake 50 HITs. One worker completed the highest number of 41 HIT assignments. We observe a typical heavy tail-like distribution in HIT completion times (mean: 37.8 secs, median: 21 secs, max: 290 secs). It is worth noting that we allocated a maximum of 5 minutes per HIT.

*Image Label Quality*
Aggregation was used to create a composite score per image given the 10 different responses for each question. We explore two different aggregation techniques. The first one is the *majority vote* where we compute the majority score given the 10 annotations for each image. The second one is the *median* method, where we compute the median across the 10 annotations for each image. Table 1 lists the summary statistics for both aggregation methods. For each aggregation technique, we compute the total number of correctly classified images for both road hazards where the consensus between the MTurk population and the field user matches. Out of 50 images which were verified via MTurk, 34 were verified as potholes, and 12 as speed-bumps, where the MTurk population and the field user labeled the image in the same category (see Table 1 and image examples in Figure 8.)

In terms of agreement with the mobile app user, 92% of images were verified with the same label as reported by the user, i.e., 46 images out of 50. Four images were labeled as ambiguous. Based on manual inspection, we found that two out of those four images contained both a pothole and a speed-bump (Figure 9a shows an example); while the remaining two images contained an unlabeled speed-bump which was not clearly visible, and hence was classified as ambiguous (Figure 9b demonstrates an example of this type).

Now we turn our focus towards assessing the reliability of annotations for hazard attributes (e.g., severity of potholes, size of speed-bumps, etc.). Please be reminded that in addition to asking users about image category, we also asked annotators to describe the attributes of the chosen hazard. To measure the inter-annotator consensus for different hazard attributes, we compute the intraclass correlation (ICC) among ratings given by the MTurk worker pool. As previously noted, our annotation procedure requires every image to be judged by $k$ annotators randomly selected from a larger population of $K$ workers ($k = 10$, while $K$ is unknown as we have no means to estimate the MTurk worker population). Consequently, $ICC(1,1)$ and $ICC(1,k)$ values, which respectively stand for single and average ICC measures [36] are computed for each of the hazard properties.

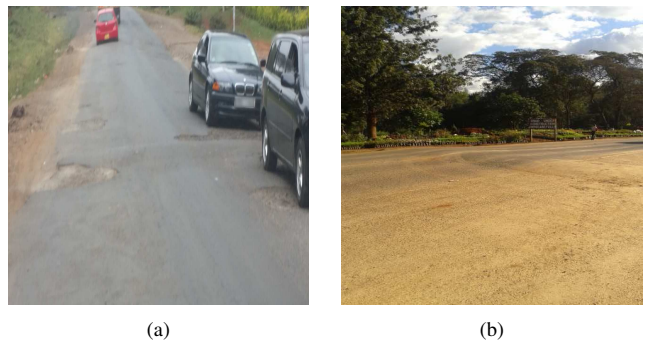

Figure 9: Misclassified images where the consensus differs between the MTurk population and the field user. Images showing faces and license plate numbers have been blurred or masked.

Table 2 reports the $ICC(1,k)$ values for all correctly verified images (i.e., 46 out of 50 images). Due to space constraints we omit $ICC(1,1)$ values. Table 2 lists the ICC values for three key hazard attributes: size and severity of potholes, and size of speed-bumps. We observe high inter-rater reliability for all hazard attributes, with all the scores being statistically significant ($p$-value $< 0.01$). Similar results were obtained for other hazard attributes. These results highlight the potential of using a crowdsourcing approach as means to verify the authenticity of the reported road hazard and its attributes.

### VISUALIZATION FRAMEWORK
Our visualization framework is a web-based application which provides a layered and an interactive (zooming and map navigation) interface, where geo-localized information from varied data sources is overlaid on top of the base map layer in an interactive fashion. It has been designed and developed using existing open-source web technologies built on top of OpenstreetMap (OSM) data.

The framework follows a layered architecture where the underlying base layer (or map layer) consists of map data from OSM, while additional layers are overlaid on top of the base layer. We visualize the location of road hazards in Nairobi in Figure 10. The location of road hazards (potholes and speed-bumps) has been contributed by our early users as part of the field experiment, as explained in previous sections.

The visualization platform is agnostic of the data source and any spatial information can be rendered as an additional layer.

|  | Min (Max) | Mean (Median) | $ICC(1,k)$ |
|---|---|---|---|
| Size of potholes | 1.0 (3.0) | 2.16 (2.0) | 0.90 |
| Severity of potholes | 1.0 (4.0) | 2.26 (2.0) | 0.91 |
| Size of speed-bumps | 1.0 (3.0) | 1.92 (2.0) | 0.73 |

Table 2: $ICC(1,k)$ scores of hazard attributes (All values are statistically significant at $p < 0.01$.) Mean and median (in brackets) values of each hazard attribute is also shown.

Additional layers can be rendered in their raw form (latitude/longitude pairs) or visualized in processed form (e.g., heatmaps), as in Figure 10. Moreover, the platform has been designed to handle large-scale datasets. The framework has been presented using Nairobi as use case but it can be easily extended for any other city, with minimal changes.

Besides the purpose of the visualization interface to provide a platform for local Nairobians to browse through the crowd-sourced data, we believe that it can serve as a platform to engage citizens, increase awareness and initiate a public dialogue on the state of road quality in Nairobi. The visualization platform is designed to give the citizen-contributed data back to the community which has helped create the data at the first place. In the process, the platform will facilitate a reliable, independent source of information about potholes and speed-bumps that can be used to alert municipal officials and allow citizens to monitor progress in resolving these hazards.

## COMPARISON WITH EXISTING SYSTEMS

As discussed in the previous sections, there exist systems which allow citizens to report civic issues (e.g., SeeClickFix [9], FixMyStreet [5], Citizens Connect [4], etc.), but none of these systems exist for Kenya. Due to the lack of any real-time traffic monitoring and broadcast systems, one of the systems which has gained popularity in Kenya and Nairobi in particular, is *ma3route* [6]. ma3route is a mobile and web platform that allows citizens to report and share information on existing traffic conditions in their city. *ma3route* publishes all user submissions on their Twitter channel [6]. As of writing, *ma3route* has more than 102K followers and has posted 178K tweets that contain in excess of 24K images and videos.

To examine the potential of social media as an alternative medium to obtain road hazard datasets, we manually coded the most recent 300 tweets from *ma3route*'s Twitter feed (most recent date: February 2 2014). We found that 45% of tweets contained information on traffic conditions and jams, 7% described road accidents, 8% of tweets reported street protests and how they were impeding the traffic flow, 2% of tweets reported road hazards, and the rest 38% discussed other topics (e.g., corruption, high fuel prices, suggestions to improve infrastructure, etc.) Out of 300 tweets, 81 (27%) of them contained an image. Only seven tweets in our sample contained information on road hazards, and out of those seven tweets, only three of them (1%) posted road hazard information with an image. Based on these findings, even though Twitter as a data collection medium looks promising, but it currently lacks the spatial coverage and topical focus offered by specialized mobile crowdsourcing. We plan to investigate the role of Twitter to collect data, as part of future work.

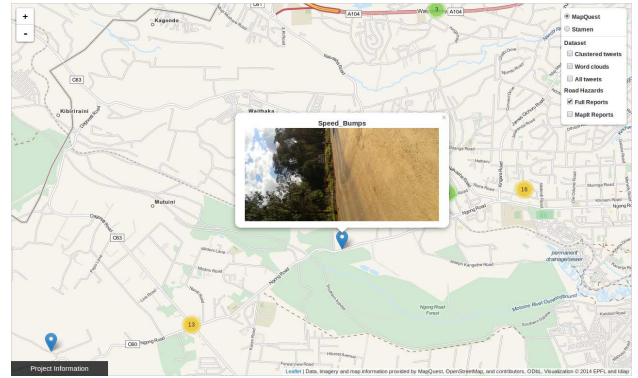

Figure 10: Visualization Framework

## DISCUSSION

In this section, we describe the technical challenges and lessons learnt while deploying *CommuniSense* in Nairobi. We further discuss *CommuniSense*'s possible role in promoting citizen engagement in Nairobi. To conclude this section, we present the implications of our findings in the design of future mobile crowdsourcing systems for the developing world.

### Technical Challenges

During the field study, we faced three major technical issues. First, due to the myriad of affordable grey market devices, we found that certain devices did not handle the mapping and location functionality well. As a result, users found it difficult to interact with the location marker on the map (Figure 4a). Second, we observed that a significant number of smartphones were still using older versions of Android (2.2 and 2.3). These versions required a different UI design, when compared to the Android version (3.0 and above) on which *CommuniSense* was developed. Although, Android provides backward compatibility, certain devices were not able to render the UI properly, causing inconvenience to users while interacting with the app. Third, for some user-submitted reports, the `base64` compression caused loss of image quality, dependent on the way the device was oriented (landscape vs. portrait mode). Additional work and experimentation is required to optimize the compression across all devices.

We used Google Play store to deploy *CommuniSense*. The processes associated with performing a limited release of the app using Google Play store, proved to be daunting for non-technical users. The process required participants to be added to a private Google Plus (G+) community. Access to G+ requires users to have a GMail account and activate alerts from their G+ profiles. Please note that only users who were part of our private G+ community, were given access to the app. Many users complained of not receiving the G+ invitation, only to discover they had not activated alerts on their G+ profiles. Even when the user successfully became part of the G+ community, they cannot search and install the app via the play store. The only way to install the app is to click *Install* on the web-interface, which then prompts the app to be installed on the device automatically (only when the device was connected to the internet). This is not the typical way users install mobile apps and so this process created confusion among early users. We believe that more work needs to be done to simplify, and streamline the process of conducting limited app release distribution via Google Play and other app distribution channels.

**Citizen Engagement**

Nairobi residents have been frustrated and lost faith in the city council's ability to improve road conditions. The words of few users highlight this sentiment:

*"Anything to do with city council would require a major overhaul of the personnel. Otherwise this would not be possible."*

*"And I do not trust that the city council would take our complaints seriously. They first need to fix the roads properly instead of patching them up year after year!"*

*"Actual or visible feedbacks would motivate me to report even paying some costs."*

*"I would only report road issues if I thought something would be done about it. I am not sure that's currently the case."*

These sentiments are shared among residents in many developing cities. We believe that the city council could benefit by leveraging the data collected by *CommuniSense*. The design of this application provides a channel to gather direct input from citizens on the condition of urban infrastructure. This would save time and money involved in manually documenting road hazards, as currently done by government engineers. The platform would also offer a mechanism to engage users into reporting hazards as well as providing accountability structures to show residents that their tax money is being used effectively.

**Relevance for Mobile Human Computer Interaction**

Crowdsourcing methods have begun to receive attention in the field of intelligent transportation systems [12, 24] and governance [17] to gather feedback from inhabitants on locally relevant issues [33]. Crowdsourcing present exciting opportunities for developing cities, as they lack monitoring technologies to obtain reliable data on urban infrastructure conditions; therefore it becomes imperative to leverage locally available resources (i.e., people) to collect such type of information. Moreover, we believe that the effectiveness of existing governance systems can be substantially enhanced by applying mobile crowdsourcing solutions, which facilitate real-time data collection, categorization, verification, and dissemination. As developing countries start looking forward towards improving social welfare and quality of life, it is important to funnel broad and meaningful feedback from community stakeholders on community needs, as well as on the effectiveness of government initiatives [17].

As a design choice, we have used smartphones to collect data. We are well aware that our smartphone-based approach presents constraints to our collection methodology in terms of reaching a much wider audience. However, recent trends in smartphone adoption and subsidized cost of smartphones demonstrate that they can be used to achieve sufficient data diversity [8, 39]. We believe the applicability of our approach is wider and generalizable to other developing cities which are facing similar problems, but the mobile platform needs to be contextualized for local needs and concerns [33, 35]. While mobile crowdsourcing has been used in the developed cities (e.g., SeeClickFix [9], FixMyStreet [5], Citizens Connect [4]), crowd-based verification which we propose in the paper, can still be applicable to these systems.

We designed, implemented and tested a mobile crowdsourcing platform for a world region that is still under-represented and under-studied in mobile HCI research. Our experience in this study can contribute to the research of mobile crowdsourcing systems in developing cities.

**CONCLUSION**

In this paper, we examined the use of mobile crowdsourcing as means to gather and document Nairobi's road quality information. First, we presented the key results of a road quality survey in Nairobi. Based on the survey's findings, we then developed a mobile crowdsourcing application, called *CommuniSense*, to collect road quality data. The application served as a tool for users to locate, describe, and photograph road hazards. We tested our application through a two-week field study amongst 30 participants to document various forms of road hazards from different areas in Nairobi. To verify the authenticity of user-contributed reports from our field study, we proposed to use online crowdsourcing using Amazon's Mechanical Turk (MTurk), to verify whether submitted reports indeed depict road hazards. We found 92% of user-submitted reports to match the MTurkers judgements. *CommuniSense* advances the research in the domain of citizen-based reporting, by integrating it with online crowd-based verification for quality control.

**ACKNOWLEDGEMENTS**

This research was partly done while D. Santani and J. Njuguna interned at IBM Research Africa. D. Santani and D. Gatica-Perez also acknowledge the support of the SNSF through the Youth@Night project. We thank anonymous reviewers and our paper shepherd for their valuable feedback.